\documentclass[twocolumn,aps,pra,showpacs,floatfix,reprint]{revtex4-1}
\usepackage{graphicx}
\usepackage{amsmath}
\usepackage{bm}
\usepackage{color}
\usepackage{dcolumn}
\usepackage{multirow}
\usepackage{hyperref}

\begin{document}

% All ``equation'' environments have been changed to ``eqnarray''
% This was necessary for ``linenumbers'' option to work

\newcommand{\etal}{{\it et al.}}
\newcommand{\fref}[1]{Fig.~\ref{#1}}
\newcommand{\Fref}[1]{Figure \ref{#1}}
\newcommand{\sref}[1]{Sec.~\ref{#1}}
\newcommand{\Eref}[1]{Eq.~(\ref{#1})}
\newcommand{\tref}[1]{Table~\ref{#1}}
\newcommand{\rtw}{\longrightarrow}
\newcommand{\veps}{\varepsilon}
\newcommand{\cm}{cm$^{-1}$}

\newcommand{\cmt}[1]{\textbf{[\![}#1\textbf{]\!]}}
\newcommand{\pprime}{{\prime\prime}}
\newcommand{\yb}[1]{$^{#1}$Yb}
\newcommand{\ls}[4]{\ensuremath{^{#1}\!{#2}_{#3}^{#4}}}
\newcommand{\jj}[4]{\ensuremath{\left(#1,#2\right)_{#3}^{#4}}}
\newcommand{\s}{\ls{1}{S}{0}{}}
\newcommand{\p}{\ls{3}{P}{0}{\circ}}
\newcommand{\clocks}{\ensuremath{6s^2~\ls{1}{S}{0}{}}}
\newcommand{\clockp}{\ensuremath{6s6p~\ls{3}{P}{0}{\circ}}}

% odd parity J=1 states
\newcommand{\odd}[1]%
{%
  \ifnum#1=1{\ensuremath{6s6p~{\ls{3}{P}{1}{\circ}}}}\fi%
  \ifnum#1=2{\ensuremath{6s6p~{\ls{1}{P}{1}{\circ}}}}\fi%
  \ifnum#1=3{\ensuremath{5d6s^2~\jj{\frac{7}{2}}{\frac{5}{2}}{1}{\circ}}}\fi%
  \ifnum#1=4{\ensuremath{5d6s^2~{^\circ_1}}}\fi%
  \ifnum#1=5{\ensuremath{6s7p~{\ls{3}{P}{1}{\circ}}}}\fi%
  \ifnum#1=6{\ensuremath{5d6s^2~{\ls{3}{D}{1}{\circ}}}}\fi%
  \ifnum#1=7{\ensuremath{6s7p~{\ls{1}{P}{1}{\circ}}}}\fi%
}

% odd parity J=1 states; resonances
\newcommand{\omegaodd}[1]%
{%
  \ifnum#1=1{0.081978}\fi%
  \ifnum#1=2{0.114219}\fi%
  \ifnum#1=3{0.131482}\fi%
  \ifnum#1=4{0.170473}\fi%
  \ifnum#1=5{0.173934}\fi%
  \ifnum#1=6{0.175065}\fi%
  \ifnum#1=7{0.184823}\fi%
}

% Affiliations
\newcommand{\NIST}{
National Institute of Standards and Technology, 
325 Broadway, Boulder, Colorado 80305, USA}

\title{Experimental constraints on the polarizabilities of the $6s^2~{\s}$ and $6s6p~{\p}$ states of Yb}

\author{K. Beloy}
\affiliation{\NIST}

\date{\today}

\begin{abstract}
We utilize accurate experimental data available in the literature to yield bounds on the polarizabilities of the ground and first excited states of atomic Yb. For the $\clocks$ ground state, we find the polarizability $\alpha$ to be constrained to $134.4<\alpha\leq144.2$ in atomic units, while for the $\clockp$ excited state we find $280.1<\alpha\leq289.9$. The uncertainty in each of these values is 1.0. These constraints provide a valuable check for {\it ab initio} and semi-empirical methods used to compute polarizabilities and other related properties in Yb.
\end{abstract}

% 06.20.Jr  Determination of fundamental constants
% 06.30.Ft 	Time and frequency
% 33.20.Bx 	Radio-frequency and microwave spectra
% 32.10.Dk 	Electric and magnetic moments, polarizabilities  
% 32.70.Cs	Oscillator strengths, lifetimes, transition moments 
% 32.60.+i	Zeeman and Stark effects  
% 31.15.ap 	Polarizabilities and other atomic and molecular properties 

\pacs{31.15.ap,32.10.Dk,32.60.+i}
\maketitle

\section{Introduction}

Lifetimes, polarizabilities, and long-range interaction parameters are intimately connected atomic properties. For a particular atomic state, these properties may be precisely determined given a complete knowledge of oscillator strengths, together with associated frequencies, for all transitions connecting to that state (we restrict our attention to transitions and interactions of the electric dipole type throughout). 
While such a complete characterization of oscillator strengths eludes experimental determination, the atomic properties themselves may be measured directly. 
Knowledge of one or more properties (e.g., lifetimes) may be exploited to obtain useful information on the other related properties (e.g., polarizabilities or van der Waals coefficients).

In this paper, we employ accurate experimental data from the literature to set upper and lower bounds on the polarizabilities of the $\clocks$ and $\clockp$ states of atomic Yb, for which experimental values are lacking. Three key parameters we utilize are the $\odd{2}$ radiative lifetime, known to better than 0.1\% from the work of Takasu~\etal~\cite{TakKomHon04etal}, the $\clocks$---$\clocks$ van der Waals coefficient $C_6$, known to better than 2\% from the work of Kitagawa~\etal~\cite{KitEnoKas08etal}, and the $\clocks$---$\clockp$ differential polarizability, known to 0.002\% from the work of Sherman~\etal~\cite{SheLemHin12etal}. 
We further supplement this data with experimental lifetime results compiled in Refs.~\cite{BlaKom94,BowBudCom96etal}. At the $1\sigma$ confidence level, our results constrain both polarizabilities to a window of width 12 atomic units. 
For the $\clocks$ polarizability, this is comparable to the best theoretical results, while for the $\clockp$ polarizability, this window is half that of the best theoretical results.

We note the similarity of this work to Ref.~\cite{DerPor02}, wherein the lifetimes of the first two excited states of atomic Cs were extracted from the measured ground state van der Waals coefficient $C_6$ and ratio of the two lifetimes.
An essential difference between Ref.~\cite{DerPor02} and the present work is that, whereas Ref.~\cite{DerPor02} included additional input from sophisticated many-body calculations, our present results are based entirely on experimental data, free from all but first-principle theoretical input.

\section{Preliminaries}

In this section, we present formulae for the ac polarizability $\alpha(\omega)$, the atom-wall interaction coefficient $C_3$, and the van der Waals coefficient $C_6$. 
The expressions here assume atoms in the ground state, with atomic units being used throughout.

The ac polarizability $\alpha(\omega)$ describes the atomic response to a harmonic electric field oscillating at (angular) frequency $\omega$. It may be written as
\begin{eqnarray}
\alpha(\omega)=\sum_i\frac{f_i}{\omega_i^2-\omega^2},
\label{Eq:pol}
\end{eqnarray}
where the summation is over all transitions from the ground state, with integration over continuum states being implicit. Here $f_i$ and $\omega_i$ represent the oscillator strength and frequency associated with the $i$-th transition, respectively ($f_i,\omega_i>0$). The {\it static} polarizability, which gives the response to a static electric field, is obtained by evaluating the ac polarizability in the zero frequency limit, $\alpha\equiv\alpha(0)$. The term polarizability, when given without specification, will imply the static polarizability.

The atom-wall interaction constant $C_3$ describes the interaction of an atom with a perfectly conductive surface. $C_3$ may be concisely expressed in terms of the ac polarizability as \cite{DzuDer10}
\begin{eqnarray*}
C_3=\frac{1}{4\pi}\int_0^\infty\alpha(i\omega)d\omega.
\end{eqnarray*}
Note that $\alpha(i\omega)$ is real; inclusion of the imaginary factor $i$ in the argument merely effects a change of sign $-\rightarrow+$ in the denominator of expression (\ref{Eq:pol}) for the ac polarizability. With Eq.~(\ref{Eq:pol}), $C_3$ may be written in terms of contributions from individual transitions,
\begin{eqnarray*}
C_3=\frac{1}{8}\sum_i\frac{f_i}{\omega_i},
\end{eqnarray*}
where the integration over $\omega$ has been performed analytically. 

The van der Waals coefficient $C_6$, which describes the long-range interaction between two atoms, may also be expressed in terms of the ac polarizability \cite{DzuDer10},
\begin{eqnarray*}
C_6=\frac{3}{\pi}\int_0^\infty|\alpha(i\omega)|^2d\omega.
\end{eqnarray*}
Using Eq.~(\ref{Eq:pol}) and performing the integration over $\omega$ analytically, we arrive at the expression
\begin{eqnarray*}
C_6=\frac{3}{2}\sum_{ij}
\frac{f_if_j}{\omega_i\omega_j\left(\omega_i+\omega_j\right)}.
\end{eqnarray*}
Note that the denominator here prohibits factorization of the summations over the indices $i$ and $j$.

Another useful property is the summation over all oscillator strengths, $\sum_i f_i$.
In the non-relativistic limit, the well-known Thomas-Reiche-Kuhn (TRK) sum rule asserts that this summation is equivalent to the total number of atomic electrons $\mathcal{N}$, i.e.,
\begin{eqnarray*}
\mathcal{N}=\sum_i f_i.
\end{eqnarray*}
Relativistic effects, however, lead to a departure from this simple interpretation; these effects will be discussed more below.

The remainder of this section is dedicated to introducing notations and conventions to be used in the following sections. We start with the definition $\alpha_i\equiv f_i/\omega_i^2$, such that $\alpha_i$ represents a partial contribution to the static polarizability from the $i$-th transition,
\begin{eqnarray*}
\alpha=\sum_i\alpha_i.
%\label{Eq:alpha}
\end{eqnarray*}
It should be noted that these contributions are positive, i.e., $\alpha_i>0$ for all $i$. Below, expressions for $C_3$, $C_6$, and $\mathcal{N}$ will be written in terms of $\alpha_i$ in favor of oscillator strengths $f_i$.

The summations over indices $i$ and $j$ in the above equations run over all allowed transitions. We will find it advantageous to partition the spectrum into a subset comprised of select lower-lying (``main'') transitions and another subset containing all remaining transitions. We use the convention of reserving indices $a$ and $b$ for the main transitions and indices $r$ and $s$ for the remaining transitions. With this convention, properties $\alpha$, $C_3$, and $\mathcal{N}$ are partitioned into two terms,
\begin{eqnarray*}
\alpha&=&\sum_a\alpha_a+\sum_r\alpha_r,\\
C_3&=&\frac{1}{8}\sum_a\alpha_a\omega_a+\frac{1}{8}\sum_r\alpha_r\omega_r,\\
\mathcal{N}&=&\sum_a\alpha_a\omega_a^2+\sum_r\alpha_r\omega_r^2.
\end{eqnarray*}
We will refer to the respective terms on each line as the ``main'' and ``tail'' terms, i.e.,
\begin{eqnarray}
\alpha&=&\alpha^\mathrm{main}+\alpha^\mathrm{tail},\nonumber\\
C_3&=&C_3^\mathrm{main}+C_3^\mathrm{tail},\nonumber\\
\mathcal{N}&=&\mathcal{N}^\mathrm{main}+\mathcal{N}^\mathrm{tail}.
\label{Eq:alphc3nsplit}
\end{eqnarray}
The $C_6$ coefficient, on the other hand, decomposes into three terms,
\begin{eqnarray*}
C_6&=&
\frac{3}{2}
\sum_{ab}
\alpha_a\alpha_b
\left(
\frac{\omega_a\omega_b}{\omega_a+\omega_b}
\right)
+\frac{3}{2}
\sum_{rs}
\alpha_r\alpha_s
\left(
\frac{\omega_r\omega_s}{\omega_r+\omega_s}
\right)
\nonumber\\&&
+3\sum_{ar}
\alpha_a\alpha_r
\left(
\frac{\omega_a\omega_r}{\omega_a+\omega_r}
\right),
\end{eqnarray*}
which we will refer to as the ``main,'' ``tail,'' and ``cross'' terms, respectively, i.e.,
\begin{eqnarray}
C_6=
C_6^\mathrm{main}
+C_6^\mathrm{tail}
+C_6^\mathrm{cross}.
\label{Eq:c6split}
\end{eqnarray}
%All terms in Eqs.~(\ref{Eq:alphc3nsplit}) and Eq.~(\ref{Eq:c6split}) are non-negative.

Finally, we introduce the frequency $\omega_0$, which we set equal to the smallest transition frequency outside of the main transitions.

\section{Limits on the polarizability}
Partitioning of the spectrum is motivated by the fact that transition frequencies and oscillator strengths for several low-lying transitions are known or can be determined sufficiently well from experimental data. With these transitions comprising the main subset, all ``main'' terms in Eqs.~(\ref{Eq:alphc3nsplit}) and Eq.~(\ref{Eq:c6split}) can be determined by directly adding the contributions from these transitions. Limited experimental information for the remaining transitions prohibits such direct evaluation of the ``tail'' and ``cross'' terms. However, given a $C_6$ coefficient and total number of electrons $\mathcal{N}$, we may set bounds on these residual terms. In particular, below we derive upper and lower bounds on $\alpha^\mathrm{tail}$ in terms of $C_6$, $\mathcal{N}$, and the various ``main'' terms. Constraints on $\alpha$ itself are then simply obtained by adding $\alpha^\mathrm{main}$ to each of these bounds.

\subsection{Upper bound on $\alpha$}

To derive an upper bound on $\alpha$, we begin by considering the factor $\omega_i\omega_j/(\omega_i+\omega_j)$, which appears in the expression for the $C_6$ coefficient. By taking partial derivatives, we find that this factor 
increases monotonically with respect to both $\omega_i$ and $\omega_j$. With this insight, we establish the following inequalities,
\begin{eqnarray*}
\frac{\omega_r\omega_s}{\omega_r+\omega_s}\geq\frac{\omega_0}{2},
\qquad\quad
\frac{\omega_a\omega_r}{\omega_a+\omega_r}\geq\frac{\omega_a\omega_0}{\omega_a+\omega_0}.
%\label{Eq:ineq1}
\end{eqnarray*}
%The first inequality from both these lines will be used to derive an upper limit on $\alpha^\mathrm{tail}$, while the second inequalities will be used to derive a lower limit on $\alpha^\mathrm{tail}$.
These inequalities may be used to give lower bounds on $C_6^\mathrm{tail}$ and $C_6^\mathrm{cross}$ in terms of $\alpha^\mathrm{tail}$,
\begin{eqnarray*}
C_6^\mathrm{tail}\geq
\frac{3}{2}\sum_{rs}\alpha_r\alpha_s\left(\frac{\omega_0}{2}\right)
=\frac{3}{4}\omega_0(\alpha^\mathrm{tail})^2,
\qquad\\
C_6^\mathrm{cross}\geq
3\sum_{ar}\alpha_a\alpha_r\left(\frac{\omega_a\omega_0}{\omega_a+\omega_0}\right)
=\frac{3}{2}\omega_0\xi\alpha^\mathrm{main}\alpha^\mathrm{tail},
\end{eqnarray*}
where we have introduced the factor $\xi$ according to the relation,
\begin{eqnarray*}
\xi\,\alpha^\mathrm{main}=\sum_a\alpha_a\frac{2\omega_a}{\omega_a+\omega_0}.
\end{eqnarray*}
$\xi$ is positive; furthermore, if $\omega_0$ is larger than all main transition frequencies, it follows that $\xi$ is necessarily less than unity as well.

The bounds on $C_6^\mathrm{tail}$ and $C_6^\mathrm{cross}$ give a corresponding bound on $C_6$ itself,
\begin{eqnarray*}
C_6\geq C_6^\mathrm{main}
+\frac{3}{4}\omega_0(\alpha^\mathrm{tail})^2
+\frac{3}{2}\omega_0\xi\alpha^\mathrm{main}\alpha^\mathrm{tail}.
\end{eqnarray*}
Noting that all factors here are necessarily non-negative, we may rearrange this inequality to yield an upper bound on $\alpha^\mathrm{tail}$,
\begin{eqnarray*}
\alpha^\mathrm{tail}\leq
-\xi\alpha^\mathrm{main}
+\sqrt{(\xi\alpha^\mathrm{main})^2+\frac{4}{3}\frac{(C_6-C_6^\mathrm{main})}{\omega_0}}.
\end{eqnarray*}
By simply adding $\alpha^\mathrm{main}$, we arrive at the result
\begin{eqnarray}
\alpha\leq
\alpha^\mathrm{main}-\xi\alpha^\mathrm{main}
+\sqrt{(\xi\alpha^\mathrm{main})^2+\frac{4}{3}\frac{(C_6-C_6^\mathrm{main})}{\omega_0}}.
\nonumber\\
\label{Eq:alphaupper}
\end{eqnarray}
%As expected, this inequality implies $\alpha=\alpha^\mathrm{main}$ in the limit of $C_6^\mathrm{main}=C_6$---that is, when {\it all} transitions are incorporated into the main subset. In the opposite limit---when no transitions are included into the main subset---we find $\alpha\leq\sqrt{(4/3)C_6/\omega_0}$, with $\omega_0$ being the first allowed transition frequency.
%
%Aside from the trivial case when all transitions are included in the main subset, the equality in (\ref{Eq:alphaupper}) only holds if all tail contributions come from transitions with a frequency precisely equal to $\omega_0$.
In the limit when no transitions are included within the main subset, we find $\alpha\leq\sqrt{(4/3)C_6/\omega_0}$, with $\omega_0$ being the first allowed transition frequency.

\subsection{Lower bound on $\alpha$}

To derive a lower bound on $\alpha$, we again start with the factor $\omega_i\omega_j/(\omega_i+\omega_j)$, this time utilizing the two inequalities
\begin{eqnarray}
\frac{\omega_r\omega_s}{\omega_r+\omega_s}\leq\frac{1}{4}\left(\omega_{r}+\omega_{s}\right),
\qquad\quad
\frac{\omega_a\omega_r}{\omega_a+\omega_r}<\omega_a.
~~\label{Eq:ineq2}
\end{eqnarray}
The first inequality follows from $4\omega_r\omega_s\leq
4\omega_r\omega_s+\left(\omega_r-\omega_s\right)^2=\left(\omega_r+\omega_s\right)^2$, while the second is obtained by taking the limit $\omega_r\rightarrow\infty$.
These inequalities may be used to set upper bounds on $C_6^\mathrm{tail}$ and $C_6^\mathrm{cross}$ in terms of $\alpha^\mathrm{tail}$,
\begin{eqnarray*}
C_6^\mathrm{tail}\leq
\frac{3}{2}\sum_{rs}\alpha_r\alpha_s\frac{1}{4}\left(\omega_{r}+\omega_{s}\right)
=6C_3^\mathrm{tail}\alpha^\mathrm{tail},\\
C_6^\mathrm{cross}\leq
3\sum_{ar}\alpha_a\alpha_r\omega_a
=24C_3^\mathrm{main}\alpha^\mathrm{tail}.
~~~
\end{eqnarray*}
The equality for $C_6^\mathrm{cross}$ here only holds if either all or none of the transitions are included in the main subset.
Together these inequalities give an upper bound on $C_6$ itself,
\begin{eqnarray}
C_6\leq C_6^\mathrm{main}
+6C_3^\mathrm{tail}\alpha^\mathrm{tail}
+24C_3^\mathrm{main}\alpha^\mathrm{tail}.
\label{Eq:notuseful}
\end{eqnarray}

In absence of direct experimental information for $C_3^\mathrm{tail}$, expression (\ref{Eq:notuseful}) is of limited utility in its present form.
However, with inspiration from elementary probability and statistics, we may find a useful limit on $C_3^\mathrm{tail}$.
Namely, for $p_r\geq0$ and $\sum_rp_r=1$, the following relations hold,
\begin{eqnarray*}
\sum_rp_r\left(x_r-\langle x\rangle\right)^2
=\langle x^2\rangle-\langle x\rangle^2\geq0,
\end{eqnarray*}
where $\langle x^n \rangle\equiv \sum_rp_rx_r^n$. In the language of probability and statistics, this merely states that variance is non-negative. By making the associations $p_r\rightarrow\alpha_r/\alpha^\mathrm{tail}$ and $x_r\rightarrow\omega_r$, we arrive at the inequality,
\begin{eqnarray*}
C_3^\mathrm{tail}\leq\frac{1}{8}\sqrt{\mathcal{N}^\mathrm{tail}\alpha^\mathrm{tail}}.
\end{eqnarray*}
Taking this result together with (\ref{Eq:notuseful}), we find
\begin{eqnarray}
C_6\leq C_6^\mathrm{main}
+\frac{3}{4}\sqrt{\mathcal{N}-\mathcal{N}^\mathrm{main}}\left(\alpha^\mathrm{tail}\right)^{3/2}
+24C_3^\mathrm{main}\alpha^\mathrm{tail},
\nonumber\\
\label{Eq:C6upper}
\end{eqnarray}
where we have further used $\mathcal{N}^\mathrm{tail}=\mathcal{N}-\mathcal{N}^\mathrm{main}$.

Our aim now is to rearrange inequality (\ref{Eq:C6upper}) to obtain bounds on $\alpha^\mathrm{tail}$. To this end, we note that in the limit of equality, expression (\ref{Eq:C6upper}) becomes a cubic equation for $\sqrt{\alpha^\mathrm{tail}}$. A general cubic equation $ax^3+bx^2+cx+d=0$ has three solutions for $x$; for the special case of $a,b\geq0$, $c=0$, and $d\leq0$---such as the present case---only one solution is a non-negative real number. This solution corresponds to a lower bound on $\sqrt{\alpha^\mathrm{tail}}$. Squaring this solution to the cubic equation and adding $\alpha^\mathrm{main}$, we find a lower bound on $\alpha$,
\begin{eqnarray}
\alpha&\geq&\alpha^\mathrm{main}+
\frac{1}{24}\frac{\left(C_6-C_6^\mathrm{main}\right)}{C_3^\mathrm{main}}
\nonumber\\&&\times
g\left(\frac{2^{14}}{9}\frac{\left(C_3^\mathrm{main}\right)^3}
{\left(C_6-C_6^\mathrm{main}\right)\left(\mathcal{N}-\mathcal{N}^\mathrm{main}\right)}\right),
\label{Eq:alphalower}
\end{eqnarray}
where the function $g(x)$ is defined for $x>0$ by
\begin{eqnarray*}
g(x)=\frac{3}{2}x\left[
2\cosh\left(\frac{1}{3}\cosh^{-1}\left(\frac{1-x}{x}\right)\right)-1
\right]^2.
\end{eqnarray*}
The function $g(x)$ increases monotonically with $x$, approaching zero [specifically, $g(x)\rightarrow3(x/2)^{1/3}$] in the limit $x\rightarrow0$ and unity in the limit $x\rightarrow\infty$. An equivalent function may be obtained by simultaneously replacing the hyperbolic cosine function and its inverse by their trigonometric counterparts (i.e., $\cosh\rightarrow\cos$, $\cosh^{-1}\rightarrow\cos^{-1}$) \cite{hypertrignote}.
In the limit of no transitions being included in the main subset, we find $\alpha\geq\left[(4/3)C_6/\sqrt{\mathcal{N}}\right]^{2/3}$.

\section{Experimental bounds on the $\clocks$ and $\clockp$ polarizabilities in Yb}

Inequalities (\ref{Eq:alphaupper}) and (\ref{Eq:alphalower}) represent the principle results of the previous section. Here we use these inequalities together with experimental data to set constraints on the $\clocks$ ground state polarizability of Yb. Moreover, we further constrain the polarizability of the $\clockp$ excited state by employing the result of a recent high-accuracy measurement of the $\clocks$---$\clockp$ {\it differential} polarizability. In the following section, these constraints are compared with theoretical values of the $\clocks$ and $\clockp$ polarizabilities given the literature.

\subsection{Bounds on the $\clocks$ polarizability}

%We start by compiling known experimental data for some low-lying transitions in Yb. 
In Table \ref{Tab:contribs}, we compile data for the lowest transitions from the ground state in Yb. Transition frequencies are taken from Ref.~\cite{NISTrareearth}, while partial contributions $\alpha_i$ are inferred from experimental lifetime data given in Refs.~\cite{BlaKom94,BowBudCom96etal,TakKomHon04etal}. We ascribe uncertainties to $\alpha_i$ based on uncertainties quoted in the relevant references. The $\clocks\rightarrow\odd{2}$ transition gives the dominant contribution to the polarizability ($\sim\!70\%$). With an accurate value for the $\odd{2}$ lifetime known from photoassociation spectroscopy~\cite{TakKomHon04etal}, this dominant contribution is determined to within 0.1\%. The $\clocks\rightarrow\odd{3}$ transition, involving an excitation from the $4f$ electronic shell, gives the next largest contribution to the polarizability ($\sim\!15\%$). We have determined this contribution by taking the weighted mean of five lifetime results compiled in Ref.~\cite{BlaKom94}. Rather than using the conventional uncertainty associated with the weighted mean, 0.7 in this case, we adopt a more conservative value of 1.2, which is commensurate with the smallest uncertainty of the five lifetime results. 

\newcolumntype{u}{D{;}{\,\pm\,}{-1}}

\begin{table}
\caption{Transition frequencies $\omega_i$ and contributions to polarizability $\alpha_i$ for the ground state of Yb. The lowest six transitions from the ground state are displayed, with $i$ encompassing all degenerate magnetic states of the upper level. Configurations given with three electrons imply excitation of an electron from the otherwise filled $4f$ shell. All values are in atomic units.
\label{Tab:contribs}}
\begin{ruledtabular}
\begin{tabular}{lcccu}
$i~~\left(\clocks\rightarrow~\right)$ 
& \hspace{10mm}
& $\omega_i$ 
&
& \multicolumn{1}{c}{$\alpha_i$} 
\\
\hline
\vspace{-2mm}\\
\odd{1} && \omegaodd{1} & 		&   2.4	; 0.1\footnotemark[1]{^,}\footnotemark[2] \\
\odd{2} && \omegaodd{2} & 		& 100.4	; 0.1\footnotemark[3] \\
\odd{3} && \omegaodd{3} & 		&  21.1	; 1.2\footnotemark[1] \\
\odd{4} && \omegaodd{4} & \multirow{3}{*}{{\Huge \}}} 	&			\\
\odd{5} && \omegaodd{5} & 		&   1.6	; 0.8\footnotemark[1]{^,}\footnotemark[2] \\
\odd{6} && \omegaodd{6} &  		&			\\
\vspace{-4mm}
\footnotetext[1]{Ref.~\cite{BlaKom94}}
\footnotetext[2]{Ref.~\cite{BowBudCom96etal}}
\footnotetext[3]{Ref.~\cite{TakKomHon04etal}}
\end{tabular}
\end{ruledtabular}
\end{table}

The last three transitions in Table \ref{Tab:contribs} have similar frequencies. The proximity of these three states to one another is cause for some special consideration. For example, based on configuration and spin multiplicity, one may suspect the first of these states to decay almost exclusively to the ground $\clocks$ state, with the second decaying almost exclusively to the $5d6s~\ls{3}{D}{1}{}$ and $6s7s~\ls{3}{S}{1}{}$ states. The validity of configuration and spin quantum numbers, however, may breakdown due to strong Coulomb mixing between these three states. While the near-degeneracy brings this additional complication, it also allows for a counterbalancing simplification by permitting us to consider the three states simultaneously. Based on experimental lifetimes from Refs.~\cite{BlaKom94,BowBudCom96etal}, we take the cumulative contribution to the polarizability from these states to be $1.6\pm0.8$; the 50\% uncertainty here is a conservative estimate of the potential effects of mixing between these states. We take a single transition frequency $\omega_i$ to be representative of these transitions; our results below are insensitive to the particular choice of $\omega_i$ when taken within the window of the three transition frequencies. 

The next transition from the ground state, beyond those shown in Table~\ref{Tab:contribs}, is the $\clocks\rightarrow\odd{7}$ transition at $\omega_i=0.184823$. Experimental lifetime results from Ref.~\cite{BlaKom94} indicate that this transition could provide a contribution as large as $\sim\!9$ to the polarizability. This would be the case if the $\odd{7}$ state decayed exclusively to the ground state (an assumption which appears to be made in Ref.~\cite{ZhaDal07}). However, electric dipole coupling to the $5d6s~\ls{1}{D}{2}{}$ and $6s7s~\ls{1}{S}{0}{}$ states may also be strong. Lacking essential information on the branching ratios of the $\odd{7}$ decay, we opt not to designate a value $\alpha_i$ for this transition. Consequently, we take the transitions in Table~\ref{Tab:contribs} to comprise our main subset, and assign to $\omega_0$ the value of the $\clocks\rightarrow\odd{7}$ transition frequency.

With $\alpha_i$ given for all transitions in the main subset (Table \ref{Tab:contribs}), the factors $\alpha^\mathrm{main}$, $\xi\alpha^\mathrm{main}$, $C_3^\mathrm{main}$, $C_6^\mathrm{main}$, and $\mathcal{N}^\mathrm{main}$ may be determined directly. To obtain bounds on $\alpha$ from inequalities (\ref{Eq:alphaupper}) and (\ref{Eq:alphalower}), we further require the van der Waals coefficient $C_6$ and total number of electrons $\mathcal{N}$. Kitagawa~\etal~\cite{KitEnoKas08etal} have probed the long-range interaction between ground state Yb atoms via photoassociation spectroscopy; in their work, the authors determined the van der Waal's coefficient to be 
\begin{eqnarray}
C_6=1932\pm30. 
\label{Eq:C6expt}
\end{eqnarray}
Taking contributions $\alpha_i$ from Table \ref{Tab:contribs} together with this $C_6$ coefficient and $\mathcal{N}=70$, we obtain the following bounds on the ground state polarizability of Yb,
\begin{eqnarray*}
134.4\pm1.0<\alpha\leq144.2\pm1.0~.
\end{eqnarray*}
The uncertainties given here were obtained by performing Monte Carlo calculations of the bounds [r.h.s.~of inequalities (\ref{Eq:alphaupper}) and (\ref{Eq:alphalower})] starting with normally distributed values of $\alpha_i$ from Table \ref{Tab:contribs} and $C_6$ from Eq.~(\ref{Eq:C6expt}). We note that the uncertainty in each bound, 1.0, is below the uncertainty of $\alpha^\mathrm{main}$ itself (see Table \ref{Tab:contribs}); this is due to covariance between the ``main'' terms which appear in inequalities (\ref{Eq:alphaupper}) and (\ref{Eq:alphalower}).

We recall that the number of electrons $\mathcal{N}$, which we used here to obtain a lower bound on $\alpha$, appeared in our expressions by invoking the TRK sum rule. The TRK sum rule, however, is only strictly valid in the nonrelativistic limit, and appreciable deviations might be suspected for a heavy atom such as Yb. The lowest order relativistic corrections scale as $(Z/c)^2$, with $Z$ being the nuclear charge ($Z=\mathcal{N}$ for a neutral atom) and $c\approx137$ being the speed of light. We find that making the substitution $\mathcal{N}\rightarrow\mathcal{N}\left[1\pm(Z/c)^2\right]=70\pm18$ in expression (\ref{Eq:alphalower}) only shifts the lower bound by $\mp0.3$, this being well within our uncertainty. Moreover, dedicated calculations have found the relativistic corrections to the TRK sum rule to be only $\approx\!1\%$ for Yb~\cite{Coh03}, suggesting that the above substitution grossly overestimates these effects. Therefore, we conclude that it is acceptable to neglect relativistic corrections to the TRK sum rule in the present case.

\subsection{Bounds on the $\clockp$ polarizability}

The atomic structure of Yb is well-suited for a frequency standard based on neutral atoms confined in an optical lattice trap \cite{KatTakPal03,PorDerFor04}.
Over the course of the last few years, Yb lattice clocks have demonstrated performance on par with the best neutral atom and single-ion frequency standards, while still holding potential for further improvement \cite{PolBarLem08etal,LemLudBar09etal}. One of the largest systematic shifts to the clock frequency is due to thermal radiation from the room temperature environment impinging upon the atomic sample~\cite{PorDer06}. In an effort to reduce the uncertainty associated with this thermal shift, Sherman~\etal~\cite{SheLemHin12etal} recently measured the static Stark shift to the $\clocks\rightarrow\clockp$ optical clock frequency to high accuracy. Their results were reported in terms of the differential polarizability,
\begin{eqnarray*}
\alpha\left(\clockp\right)-\alpha\left(\clocks\right)=145.726\pm0.003~.
\end{eqnarray*}
We use this result, together with our results above for the ground state polarizability, to set constraints on the polarizability of the $\clockp$ excited state,
\begin{eqnarray*}
280.1\pm1.0<\alpha\leq289.9\pm1.0~.
\end{eqnarray*}
The excited state polarizability is about twice as large as the ground state polarizability; as a consequence, the constraint on the excited state polarizability is fractionally about a factor of two more narrow than the constraint on the ground state polarizability.

\section{Comparison with theoretical values from the literature}

In Figure \ref{Fig:alphacompare}, we display results of {\it ab initio} and semi-empirical calculations taken from the literature for the $\clocks$ and $\clockp$ polarizabilities. Along with these values, we further display the present constraints derived from experimental data.

\newcommand{\fcite}[2]{\parbox[t]{1mm}{$^\mathrm{#1}$}
Ref.~\cite{#2}}

\begin{figure}
\includegraphics[scale=0.72]{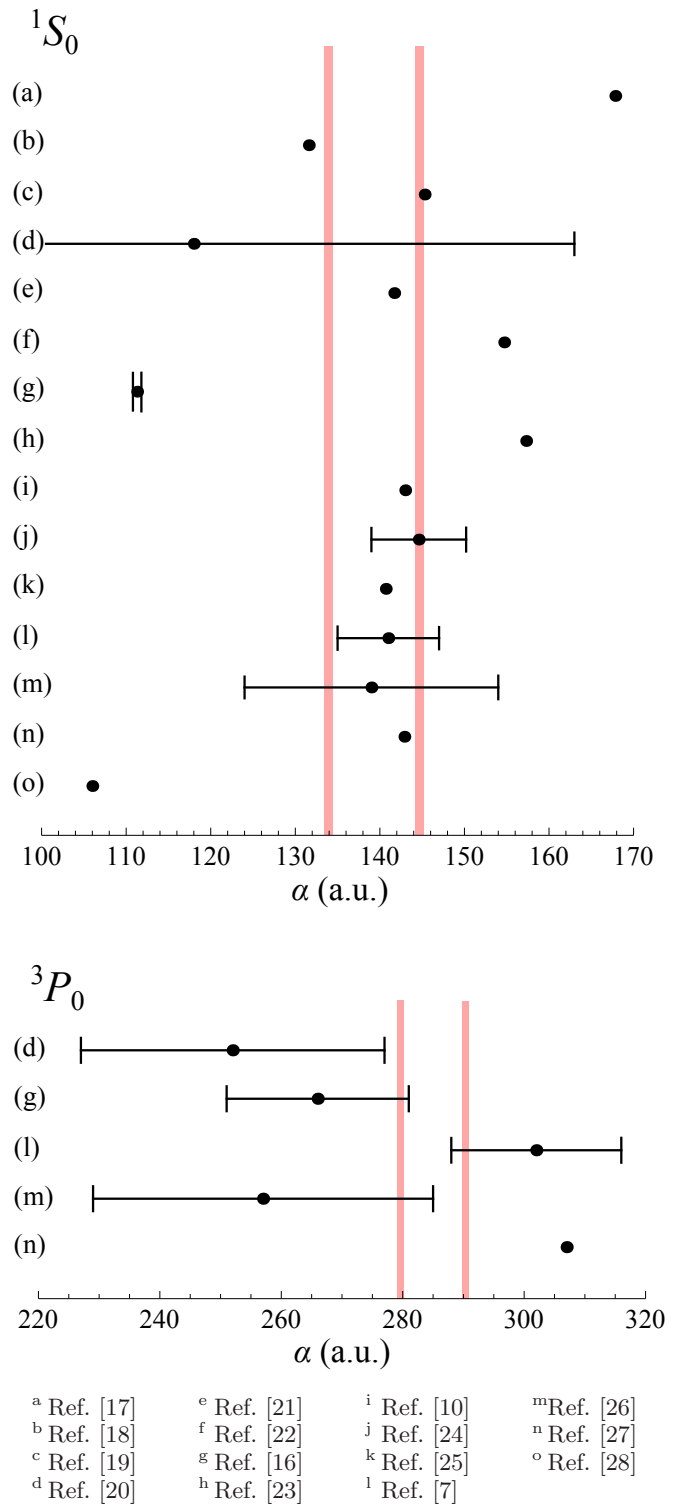}
% All the minipages are so that references compile in the desired order
\vspace{-1mm}\\
\begin{minipage}[b]{15mm}
\begin{flushleft}
\fcite{a}{KelRemSab95}\\
\fcite{b}{WanPanSch95}\\
\fcite{c}{WanDol98}\\
\fcite{d}{PorRakKoz99}
\end{flushleft}
\end{minipage}
\hspace{5mm}
\begin{minipage}[b]{15mm}
\begin{flushleft}
\fcite{e}{Mil02}\\
\fcite{f}{BucSzcCha06}\\
\fcite{g}{PorDer06}\\
\fcite{h}{ChuDalGro07}
\end{flushleft}
\end{minipage}
\hspace{5mm}
\begin{minipage}[b]{15mm}
\begin{flushleft}
\fcite{i}{ZhaDal07}\\
\fcite{j}{SahDas08}\\
\fcite{k}{ThiSch09}\\
\fcite{l}{DzuDer10}
\end{flushleft}
\end{minipage}
\hspace{5mm}
\begin{minipage}[b]{15mm}
\begin{flushleft}
\fcite{m}{GuoWanYe10}\\
\fcite{n}{Buc11}\\
\fcite{o}{BosFurBra11}\\
~
\end{flushleft}
\end{minipage}
\caption{(color online) Theoretical values from the literature for the polarizabilities of the $\clocks$ (top panel) and $\clockp$ (bottom panel) states of Yb, together with the present constraints derived from experimental data. Theoretical values are denoted with circle markers, along with error bars where available. The vertical bands indicate the present constraints, with the left band extending 1$\sigma$ below the lower bound and right band extending 1$\sigma$ above the upper bound.}
\label{Fig:alphacompare}
\end{figure}

For the ground state, we find that several of the theoretical values lie within our constraints. In particular, with the exception of Ref.~\cite{PorDer06}, all results with explicit error bars are found to agree very well with our constraints. Dzuba and Derevianko \cite{DzuDer10} identified an oversight in the semi-empirical method of Ref.~\cite{PorDer06}, which accounts for the discrepancy with this result. At the $1\sigma$ confidence level, we see that our present results and the most accurate theoretical results of Refs.~\cite{SahDas08,DzuDer10} constrain the polarizability to a similar window, having a width of about 12.

For the excited state, we find that the theoretical results with explicit error bars agree reasonably well with our constraints. Specifically, values from  Refs.~\cite{PorRakKoz99,PorDer06,GuoWanYe10} are seen to lie about $1\sigma$ below our lower bound, while the value from Ref.~\cite{DzuDer10} lies about $1\sigma$ above our upper bound. The only other result, having no explicit error bar, also lies above our upper bound. At the $1\sigma$ confidence level, we see that our results constrain the polarizability to a tighter window---about a factor of two narrower---than the best theoretical results.

It is not the goal of the present paper to provide criticism of the theoretical methods used in 
Refs.~\cite{KelRemSab95,WanPanSch95,WanDol98,
PorRakKoz99,Mil02,BucSzcCha06,PorDer06,ChuDalGro07,
ZhaDal07,SahDas08,ThiSch09,DzuDer10,GuoWanYe10,Buc11,BosFurBra11}. 
Nevertheless, as a general comment, it appears that our constraints give a certain degree of validation for several of these methods. We anticipate that our results will serve as a valuable check for {\it ab initio} and semi-empirical methods used in future works to calculate polarizabilities, as well as other related properties of Yb (e.g., $C_3$ and $C_6$ coefficients, magic wavelengths, dipole matrix elements). Fig.~\ref{Fig:alphacompare} illustrates that our present constraints are relevant for this purpose.

\section{Comments on the spectral distribution of $\alpha^\mathrm{tail}$}

The gap between our upper and lower bounds demonstrates our ignorance in the spectral distribution of $\alpha^\mathrm{tail}$ across the frequencies $\omega_r\geq\omega_0$. For example, the equality of (\ref{Eq:alphaupper}), which gives the upper bound on the polarizability, is only satisfied if the entirety of $\alpha^\mathrm{tail}$ is accumulated from transitions with frequency precisely $\omega_0$ and no larger. With an appropriate model for the distribution of  $\alpha^\mathrm{tail}$ (or, equivalently, distribution of oscillator strengths), we could arrive at a central value for the polarizability lying somewhere between our present constraints. We have chosen not pursue such a program, stressing that our present results are derived from experimental data available in the literature and are independent of such theoretical modeling.

Unlike upper bound (\ref{Eq:alphaupper}), the lower bound (\ref{Eq:alphalower}) does not correspond to a precise distribution of $\alpha^\mathrm{tail}$. From numerical calculations, we found that, with $C_6$, $\mathcal{N}$, and ``main'' terms fixed, $\alpha^\mathrm{tail}$ is minimized by a distribution peaked sharply about a frequency $\omega_r\approx2.5$.
This realization motivated the choice of inequalities in (\ref{Eq:ineq2}). The first is near equality when $|\omega_r-\omega_s|\ll(\omega_r+\omega_s)$, while the second is near equality for $\omega_a\ll\omega_r$. Therefore, with these inequalities our analytical lower bound (\ref{Eq:alphalower}) has a close correspondence to this minimal distribution.
%, though yields a value slightly below the minimal $\alpha^\mathrm{tail}$.
While it is physically unlikely that $\alpha^\mathrm{tail}$ is accumulated entirely at frequency $\omega_r\approx2.5$, we again emphasize that we do not speculate on the actual spectral distribution of $\alpha^\mathrm{tail}$.

\section{Extension to non-static polarizabilities}
While the primary focus of this work is to obtain constraints on the static polarizability $\alpha\equiv\alpha(0)$, the results above can further be extended to yield constraints on the ac polarizability $\alpha(\omega)$ evaluated at other frequencies $\omega\neq0$.
To illustrate this, we consider the ac polarizability evaluated at the ``magic'' lattice frequency $\omega^*$, which balances the ac polarizabilities of the \clocks{} and \clockp{} clock states \cite{KatTakPal03}. Dedicated measurements have determined the magic frequency to be $\omega^*=0.06000$ \cite{LemLudBar09etal,BarStaLem08etal}. Defining $\beta\equiv\alpha(\omega^*)$, it follows from Eq.~(\ref{Eq:pol}) that contributions to $\beta$ satisfy
\begin{eqnarray*}
\beta_i=\alpha_i\frac{\omega_i^2}{\omega_i^2-\omega^{*2}}.
\end{eqnarray*}
The ``main'' term $\beta^\mathrm{main}\equiv\sum_a\beta_a$ can be computed directly from the values given in Table \ref{Tab:contribs}, while the ``tail'' term $\beta^\mathrm{tail}\equiv\sum_r\beta_r$ is necessarily limited to the range
\begin{eqnarray}
\alpha^\mathrm{tail}<\beta^\mathrm{tail}\leq\alpha^\mathrm{tail}\left(\frac{\omega_0^2}{\omega_0^2-\omega^{*2}}\right),
\label{Eq:betatail}
\end{eqnarray}
where the factor in parenthesis evaluates to 1.12. Using inequalities (\ref{Eq:betatail}) together with inequalities (\ref{Eq:alphaupper}) and (\ref{Eq:alphalower}) for bounds on $\alpha^\mathrm{tail}$, we obtain the following constraints on $\beta=\beta^\mathrm{main}+\beta^\mathrm{tail}$,
\begin{eqnarray*}
181.1\pm1.3<\beta\leq 193.2\pm1.1~.
\end{eqnarray*}
These constraints have been derived specifically for the \clocks{} ground state; however, we note that they are equally applicable for the \clockp{} excited state, following from the definition of magic frequency, i.e.~%
$\beta\left(\clockp\right)=\beta\left(\clocks\right)$. Knowledge of $\beta\equiv\alpha(\omega^*)$ is useful, as it can be used to directly relate lattice intensity to trap depth in the optical lattice clock \cite{BarStaLem08etal}.

\section{Conclusion}

We have utilized experimental data available in the literature to yield bounds on the polarizabilities of the $\clocks$ and $\clockp$ states of Yb.  
Key experimental parameters employed were the $\odd{2}$ radiative lifetime from Takasu~\etal~\cite{TakKomHon04etal}, the $\clocks$---$\clocks$ van der Waals coefficient $C_6$ from Kitagawa~\etal~\cite{KitEnoKas08etal}, and the $\clocks$---$\clockp$ differential polarizability from Sherman~\etal~\cite{SheLemHin12etal}.
For the $\clocks$ state, our results constrain the polarizability to a window comparable to that of the best theoretical results (at the $1\sigma$ confidence level), while for the $\clockp$ state, our results constrain the polarizability to a window half the width of the best theoretical results.
We anticipate that our results will serve as a valuable check for {\it ab initio} and semi-empirical methods aimed at calculating polarizabilities and other related properties in Yb.

\begin{acknowledgements}
This work was funded through the Research Associateship Programs of the National Research Council. The author thanks J.~A.~Sherman, A.~D.~Ludlow, A.~Derevianko, and C.~W.~Oates for helpful discussions.
\end{acknowledgements}

%\bibliography{Yb_pol_limits_biblio}

\begin{thebibliography}{29}
\bibitem [{hyp()}]{hypertrignote}%
  \BibitemOpen
  \href@noop {} {}\bibinfo {note} {{ We specify that for $|u|\leq1$,
  $\cos^{-1}u$ is real and satisfies $0\leq\cos^{-1}u\leq\pi$, while for
  $u\geq1$, $\cosh^{-1}u$ is real and satisfies $\cosh^{-1}u\geq0$. We further
  take the relation $\cosh^{-1}u=\pm i\cos^{-1}u$ to hold, with the choice of
  sign being inconsequential. }}\BibitemShut {NoStop}%
\end{thebibliography}

\end{document}